# A New Approach to Design Graph Based Search Engine for Multiple Domains Using Different Ontologies

Debajyoti Mukhopadhyay, Sukanta Sinha

Web Intelligence & Distributed Computing Research Lab, Techno India Group
West Bengal University of Technology
EM 4/1, Salt Lake Sector V, Calcutta 700091, India
{debajyoti.mukhopadhyay, sukantasinha2003}@gmail.com

*Abstract* − Search Engine has become a major tool for searching any information from the World Wide Web (WWW). While searching the huge digital library available in the WWW, every effort is made to retrieve the most relevant results. But in WWW majority of the Web pages are in HTML format and there are no such tags which tells the crawler to find any specific domain. To find more relevant result we use Ontology for that particular domain. If we are working with multiple domains then we use multiple ontologies. Now in order to design a domain specific search engine for multiple domains, crawler must crawl through the domain specific Web pages in the WWW according to the predefined ontologies.

## 1. Introduction

Internet is an infinite reservoir of information. It has bought the concept of *Vasudeva Kutumbakam* to reality. To find information from the internet we needs a document retrieval system called search engine. A Web search engine mainly searches for the documents in the WWW. A Web crawler is a program that crawls through the WWW and returns the Web pages in its way, to search engine. After getting a predefined number of Web pages the crawler stops running. The search engine allows one to ask for content meeting specific criteria (typically those containing a given word or phrase). And searches the given word or phrase in the Web pages returned by the Web crawler. Then it retrieves a list of items that match those criteria. And produce a ranked list of URLs in which the keywords matched. Although such technologies are mostly used, users are still often faced with the daunting task of sifting through multiple pages of results, many of which are irrelevant.

In this paper, we discuss the basic idea of the graph based searching and describe a design and development methodology for multiple domain specific search engine based on multiple ontology matching and relevance limits which not only overcomes the problem of knowledge overhead but also supports conventional queries. Further, it is able to produce exact answer from the graph that satisfies user queries.

## 2. Domain Specific Web Search Crawling

In this section we describe working principle of a single domain specific crawler and multiple domains specific crawler.

### 2.1 Single Domain Specific Crawler

In domain specific Web search crawler, the crawler crawls down the pages which are relevant to our domain. To find the domain we need to visit all the Web pages and calculate the relevance value. Now the situation such like that the page is not related to the given domain but it belongs to another domain. For this we want to give a new proposal to working with the multiple domains. In Figure 1 we show the single domain specific crawler crawling activity.

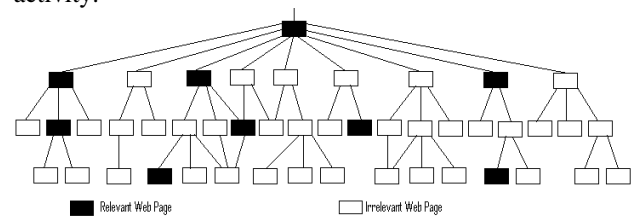

**Fig.1.** Single Domain Specific Crawling

### 2.2 Multiple Domains Specific Crawler

In multiple domains specific Web search crawler crawls down the Web pages and checking multiple domains simultaneously by using multiple Ontology terms and finding which page is related to which domain. The Web page not only related to the single domain but also it may be related with multiple domains. In our approach we taking a track to finding a Web page related to how many domains and what are their relevance scores. In Figure 2 shown how we are working with multiple domains.

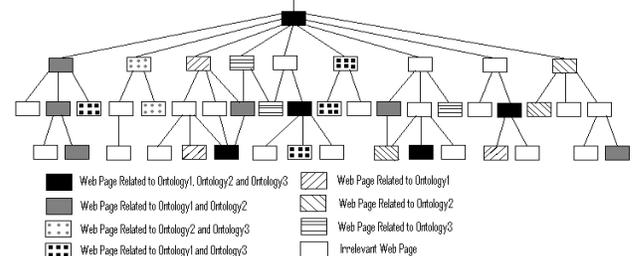

**Fig.2.** Multiple Domains Specific Crawling

## 3. Ontology Based Domain Specific Crawling

In this section we will describe what is ontology and how ontology can be used to domain specific crawling.

### 3.1 Introduction to Ontologies
The term *ontology* is a data model that represents a set of concepts within a domain and the relationships between those concepts. It is used to reason about the objects within that domain. Ontologies are used in artificial intelligence, the Semantic Web, software engineering, biomedical informatics, Library Science, and information architecture as a form of knowledge representation about the world or some part of it. Ontology is a formal description of concepts and the relationships between them. Definitions associate the names of entities in the ontology with human-readable text that describes what the names mean. The ontology can also contain rules that constrain the interpretation and use of these terms.

Ontology can be used to define common vocabularies for users who want to share knowledge about a domain. It includes definitions of concepts and relations between them, and is written in a language that can also be interpreted by a computer. Ontologies can be used to share common understanding of the structure of information, enable reuse of domain knowledge, separate domain knowledge from operational knowledge and analyze domain knowledge. [4]

### 3.2 Ontology Based Crawling
Ontology can be used in domain specific crawlers. A domain specific crawler uses ontology to describe the area of interest, in the same way as a search in a search engine uses a list of keywords to describe the area of interest. A problem with standard keyword based search queries is that it is difficult to express advanced search queries. By using ontology it is possible to express richer and more accurate queries. The system has an ontology that describes the area in which the search will be performed, and the user enters different parameters to say what should be weighted in the search. Then the program crawls the Web for pages containing text that describes the area given by the ontology.

## 4. WordNet

WordNet [6] is a semantic lexicon for the English language. A *semantic lexicon* is a dictionary of words labeled with semantic classes so associations can be drawn between words that have not previously been encountered. WordNet groups' English words into sets of synonyms called *synsets*, provides short, general definitions, and records the various semantic relations between these synonym sets. The purpose is twofold: to produce a combination of dictionary and *thesaurus* that is more intuitively usable, and to support automatic text analysis and artificial intelligence applications. A thesaurus is an indexed compilation of words with similar, related and opposite meanings.

Syntable is one type of table which contains synonyms of all ontology terms in a table. We generate a syntable for each ontology to generate more accurate relevance score of a Web page.

## 5. Proposed Approach

In our approach we crawl through the Web and add Web pages to the database which are related to our specified domains (i.e. related to our specified ontologies) and discard Web pages which are not related to our domains. To finding the domains in which a Web page belongs to or not we calculate relevance scores of that Web page for all domains, and if the relevance scores of the page are more than predefined scores then we say that the page is specific to these domains under consideration. In this section we will show the process of relevance calculation using multiple ontologies as described in section 3.1 and how this process can be used to determine whether a page is related to our specified domains or not.

### 5.1 Relevance Calculation for All Domains
In this section we describe our own algorithm depending on which we calculate relevancy of a Web page for multiple domains. Our algorithm is very simple and very effective as well. Here we assign some weight to all the ontology terms and use these weights for relevance calculation for each domain.

#### 5.1.1 Weight Table
Weight table is a table which is constructed using the given ontologies. This table contains two columns; one column for the ontology terms and another for term corresponding weights. For a term which belongs to more than one domain we assign different weights according strength to their respective domains. The strategy of assigning weights is that, the more specific term will have more weight on it. And the terms which are common to more than one domain have less weight. The sample Weight table for some terms of a given ontology of the table shown below:

| Term | Weight |
|---|---|
| Wicket | 1.0 |
| Bat | 1.0 |
| Crease | 0.8 |
| Test Match | 0.6 |
| One day Match | 0.4 |
| Ball | 0.2 |
| Ground | 0.1 |
| Player | 0.1 |

**Fig.3.** Weight table for some terms in cricket ontology

### 5.1.2 Relevance calculation algorithm

In this section we design an algorithm which calculates relevance scores of a Web page for multiple domains. Each domain represents by an ontology. Now we are making a weight table for each ontology and also making syntable which contains all the synonyms of the ontology terms by using WordNet. WordNet help us to giving more prominent result. Here we are taking a Web page, weight tables for each domain ontology terms and syntables for each domain ontology terms as input. And the following algorithm calculates relevance scores for each domain.

**INPUT:** A Web page ($\sigma$), Weight Tables for each domain and SynTables for each domain.
**OUTPUT:** The Relevance score of the Web page ($\sigma$) for each domain.

**Step1** Initialize the relevance scores of the Web page ($\sigma$) for each domain to 0. REL_ONT$_1$_$\sigma$ = 0, REL_ONT$_2$_$\sigma$ = 0 and REL_ONT$_3$_$\sigma$ = 0.

**Step2** Select first common ontology term ($\alpha$) for all domains (Ontology$_1$, Ontology$_2$ and Ontology$_3$) and corresponding weights ($w_1$, $w_2$ and $w_3$) from the weight tables.

**Step3** Calculate how many times the term ($\alpha$) occurs in the Web page ($\sigma$). Let the number of occurrence is calculated in COUNT.

**Step4** Multiply the number of occurrence calculated at Step3 with the weights $w_1$, $w_2$ and $w_3$. Let call these TERM_WEIGHT$_1$, TERM_WEIGHT$_2$ and TERM_WEIGHT$_3$. Now TERM_WEIGHT$_1$ = COUNT * $w_1$, TERM_WEIGHT$_2$ = COUNT * $w_2$, TERM_WEIGHT$_3$ = COUNT * $w_3$.

**Step5** Add these term weights to REL_ONT$_1$_$\sigma$, REL_ONT$_2$_$\sigma$ and REL_ONT$_3$_$\sigma$. So updated value of Relevancy of that Web page ($\sigma$) for the different domains are
REL_ONT$_1$_$\sigma$ = REL_ONT$_1$_$\sigma$ + TERM_WEIGHT$_1$,
REL_ONT$_2$_$\sigma$ = REL_ONT$_2$_$\sigma$ + TERM_WEIGHT$_2$,
REL_ONT$_3$_$\sigma$ = REL_ONT$_3$_$\sigma$ + TERM_WEIGHT$_3$.

**Step6** Select the corresponding SynTerm s$\alpha_1$, s$\alpha_2$ and s$\alpha_3$ from SynTables of Ontology$_1$, Ontology$_2$ and Ontology$_3$ respectively for the common term ($\alpha$).

**Step7** Calculate how many times the terms s$\alpha_1$, s$\alpha_2$ and s$\alpha_3$ occurs in the Web page ($\sigma$). Let the number of occurrence are SCOUNT$_1$, SCOUNT$_2$ and SCOUNT$_3$. IF s$\alpha_1$, s$\alpha_2$ and s$\alpha_3$ are not exists then Set SCOUNT$_1$ = 0, SCOUNT$_2$ = 0 and SCOUNT$_3$ = 0.

**Step8** Multiply the number of occurrence calculated at Step7 with the weights $w_1$, $w_2$ and $w_3$. Now TERM_WEIGHT$_1$ = SCOUNT$_1$ * $w_1$, TERM_WEIGHT$_2$ = SCOUNT$_2$ * $w_2$, TERM_WEIGHT$_3$ = SCOUNT$_3$ * $w_3$.

**Step9** Add these term weights to REL_ONT$_1$_$\sigma$, REL_ONT$_2$_$\sigma$ and REL_ONT$_3$_$\sigma$. So updated value of Relevancy of that Web page ($\sigma$) for the different domains are
REL_ONT$_1$_$\sigma$ = REL_ONT$_1$_$\sigma$ + TERM_WEIGHT$_1$,
REL_ONT$_2$_$\sigma$ = REL_ONT$_2$_$\sigma$ + TERM_WEIGHT$_2$,
REL_ONT$_3$_$\sigma$ = REL_ONT$_3$_$\sigma$ + TERM_WEIGHT$_3$.

**Step10** Select the next SynTerms from SynTables of Ontology$_1$, Ontology$_2$ and Ontology$_3$ respectively for the common term ($\alpha$) and go to Step7 until all the SynTerms from SynTables for the common term ($\alpha$) are visited.

**Step11** Select the next common term for all domains and weights from weight tables and go to Step3 until all the common terms for all domains are visited.

**Step12** Select ontology term ($\beta_1$) from the remaining ontology terms and $\beta_1$ exists in both Ontology$_1$ and Ontology$_2$ and corresponding weights ($w_1$ and $w_2$) from the weight tables.

**Step13** Calculate how many times the term ($\beta_1$) occurs in the Web page ($\sigma$). Let the number of occurrence is calculated in COUNT.

**Step14** Multiply the number of occurrence calculated at Step13 with the weights $w_1$ and $w_2$. Now TERM_WEIGHT$_1$ = COUNT * $w_1$ and TERM_WEIGHT$_2$ = COUNT * $w_2$. **Step15** Add these term weights to REL_ONT$_1$_$\sigma$ and REL_ONT$_2$_$\sigma$. So updated value of Relevancy of that Web page ($\sigma$) for the different domains are
REL_ONT$_1$_$\sigma$ = REL_ONT$_1$_$\sigma$ + TERM_WEIGHT$_1$,
REL_ONT$_2$_$\sigma$ = REL_ONT$_2$_$\sigma$ + TERM_WEIGHT$_2$.

**Step16** Select the corresponding SynTerm s$\beta_{11}$ and s$\beta_{12}$ from SynTables of Ontology$_1$, and Ontology$_2$ respectively for the common term ($\beta_1$).

**Step17** Calculate how many times the terms s$\beta_{11}$ and s$\beta_{12}$ occurs in the Web page ($\sigma$). Let the number of occurrence are SCOUNT$_1$ and SCOUNT$_2$. IF s$\beta_{11}$ and s$\beta_{12}$ are not exists then Set SCOUNT$_1$ = 0 and SCOUNT$_2$ = 0.

**Step18** Multiply the number of occurrence calculated at Step17 with the weights $w_1$ and $w_2$. Now TERM_WEIGHT$_1$ = SCOUNT$_1$ * $w_1$ and TERM_WEIGHT$_2$ = SCOUNT$_2$ * $w_2$. **Step19** Add these term weights to REL_ONT$_1$_$\sigma$ and REL_ONT$_2$_$\sigma$. So updated value of Relevancy of that Web page ($\sigma$) for the different domains are
REL_ONT$_1$_$\sigma$ = REL_ONT$_1$_$\sigma$ + TERM_WEIGHT$_1$ and
REL_ONT$_2$_$\sigma$ = REL_ONT$_2$_$\sigma$ + TERM_WEIGHT$_2$.

**Step20** Select the next SynTerms from SynTables of Ontology$_1$ and Ontology$_2$ respectively for the common term ($\beta_1$) and go to Step17 until all the SynTerms from SynTables for the common term ($\beta_1$) is visited.

**Step21** Select the next common term for Ontology$_1$ and Ontology$_2$ and weights from weight tables and go to Step13 until all the common terms for $Ontology_1$ and $Ontology_2$ are visited.

**Step22** Select ontology term ($β_2$) from the remaining ontology terms and $β_2$ exists in both $Ontology_2$ and $Ontology_3$ and corresponding weights ($w_2$ and $w_3$) from the weight tables.

**Step23** Calculate how many times the term ($β_2$) occurs in the Web page ($σ$). Let the number of occurrence is calculated in COUNT.

**Step24** Multiply the number of occurrence calculated at Step23 with the weights $w_2$ and $w_3$. Now TERM_WEIGHT$_2$ = COUNT * $w_2$ and TERM_WEIGHT$_3$ = COUNT * $w_3$. **Step25** Add these term weights to REL_ONT$_2$_$σ$ and REL_ONT$_3$_$σ$. So updated value of Relevancy of that Web page ($σ$) for the different domains are REL_ONT$_2$_$σ$ = REL_ONT$_2$_$σ$ + TERM_WEIGHT$_2$, REL_ONT$_3$_$σ$ = REL_ONT$_3$_$σ$ + TERM_WEIGHT$_3$.

**Step26** Select the corresponding SynTerm $sβ_{22}$ and $sβ_{23}$ from SynTables of $Ontology_2$, and $Ontology_3$ respectively for the common term ($β_2$).

**Step27** Calculate how many times the terms $sβ_{22}$ and $sβ_{23}$ occurs in the Web page ($σ$). Let the number of occurrence are SCOUNT$_2$ and SCOUNT$_3$. IF $sβ_{22}$ and $sβ_{23}$ are not exists then Set SCOUNT$_2$ = 0 and SCOUNT$_3$ = 0.

**Step28** Multiply the number of occurrence calculated at Step27 with the weights $w_2$ and $w_3$. Now TERM_WEIGHT$_2$ = SCOUNT$_2$ * $w_2$ and TERM_WEIGHT$_3$ = SCOUNT$_3$ * $w_3$. **Step29** Add these term weights to REL_ONT$_2$_$σ$ and REL_ONT$_3$_$σ$. So updated value of Relevancy of that Web page ($σ$) for the different domains are REL_ONT$_2$_$σ$ = REL_ONT$_2$_$σ$ + TERM_WEIGHT$_2$ and REL_ONT$_3$_$σ$ = REL_ONT$_3$_$σ$ + TERM_WEIGHT$_3$.

**Step30** Select the next SynTerms from SynTables of $Ontology_2$ and $Ontology_3$ respectively for the common term ($β_2$) and go to Step27 until all the SynTerms from SynTables for the common term ($β_2$) is visited.

**Step31** Select the next common term for $Ontology_2$ and $Ontology_3$ and weights from weight tables and go to Step23 until all the common terms for $Ontology_2$ and $Ontology_3$ are visited.

**Step32** Select ontology term ($β_3$) from the remaining ontology terms and $β_3$ exists in both $Ontology_1$ and $Ontology_3$ and corresponding weights ($w_1$ and $w_3$) from the weight tables.

**Step33** Calculate how many times the term ($β_3$) occurs in the Web page ($σ$). Let the number of occurrence is calculated in COUNT.

**Step34** Multiply the number of occurrence calculated at Step33 with the weights $w_1$ and $w_3$. Now TERM_WEIGHT$_1$ = COUNT * $w_1$ and TERM_WEIGHT$_3$ = COUNT * $w_3$. **Step35** Add these term weights to REL_ONT$_1$_$σ$ and REL_ONT$_3$_$σ$. So updated value of Relevancy of that Web page ($σ$) for the different domains are REL_ONT$_1$_$σ$ = REL_ONT$_1$_$σ$ + TERM_WEIGHT$_1$, REL_ONT$_3$_$σ$ = REL_ONT$_3$_$σ$ + TERM_WEIGHT$_3$.

**Step36** Select the corresponding SynTerm $sβ_{31}$ and $sβ_{33}$ from SynTables of $Ontology_1$, and $Ontology_3$ respectively for the common term ($β_3$).

**Step37** Calculate how many times the terms $sβ_{31}$ and $sβ_{33}$ occurs in the Web page ($σ$). Let the number of occurrence are SCOUNT$_1$ and SCOUNT$_3$. IF $sβ_{31}$ and $sβ_{33}$ are not exists then Set SCOUNT$_1$ = 0 and SCOUNT$_3$ = 0.

**Step38** Multiply the number of occurrence calculated at Step37 with the weights $w_1$ and $w_3$. Now TERM_WEIGHT$_1$ = SCOUNT$_1$ * $w_1$ and TERM_WEIGHT$_3$ = SCOUNT$_3$ * $w_3$. **Step39** Add these term weights to REL_ONT$_1$_$σ$ and REL_ONT$_3$_$σ$. So updated value of Relevancy of that Web page ($σ$) for the different domains are REL_ONT$_1$_$σ$ = REL_ONT$_1$_$σ$ + TERM_WEIGHT$_1$ and REL_ONT$_3$_$σ$ = REL_ONT$_3$_$σ$ + TERM_WEIGHT$_3$.

**Step40** Select the next SynTerms from SynTables of $Ontology_1$ and $Ontology_3$ respectively for the common term ($β_3$) and go to Step37 until all the SynTerms from SynTables for the common term ($β_3$) is visited.

**Step41** Select the next common term for $Ontology_1$ and $Ontology_3$ and weights from weight tables and go to Step33 until all the common terms for $Ontology_1$ and $Ontology_3$ are visited.

**Step42** Select remaining ontology terms $γ_1$, $γ_2$ and $γ_3$ for $Ontology_1$, $Ontology_2$ and $Ontology_3$ respectively and corresponding weights $w_1$, $w_2$ and $w_3$ from the weight tables.

**Step43** Calculate how many times the terms $γ_1$, $γ_2$ and $γ_3$ occurs in the Web page ($σ$). Let the number of occurrence are calculated in COUNT$_1$, COUNT$_2$ and COUNT$_3$.

**Step44** Multiply the number of occurrence calculated at Step43 with the weights $w_1$, $w_2$ and $w_3$. Now TERM_WEIGHT$_1$ = COUNT$_1$ * $w_1$, TERM_WEIGHT$_2$ = COUNT$_2$ * $w_2$, TERM_WEIGHT$_3$ = COUNT$_3$ * $w_3$.

**Step45** Add these term weights to REL_ONT$_1$_$σ$, REL_ONT$_2$_$σ$ and REL_ONT$_3$_$σ$. So updated value of Relevancy of that Web page ($σ$) for the different domains are REL_ONT$_1$_$σ$ = REL_ONT$_1$_$σ$ + TERM_WEIGHT$_1$, REL_ONT$_2$_$σ$ = REL_ONT$_2$_$σ$ + TERM_WEIGHT$_2$, REL_ONT$_3$_$σ$ = REL_ONT$_3$_$σ$ + TERM_WEIGHT$_3$.

**Step46** Select the corresponding SynTerm $sγ_{11}$, $sγ_{22}$ and $sγ_{33}$ from SynTables of $Ontology_1$, $Ontology_2$ and $Ontology_3$ respectively for $γ_1$, $γ_2$ and $γ_3$.

**Step47** Calculate how many times the terms $sγ_{11}$, $sγ_{22}$ and $sγ_{33}$ occurs in the Web page ($σ$). Let the number of occurrence are SCOUNT$_1$, SCOUNT$_2$ and SCOUNT$_3$. IF $sγ_{11}$, $sγ_{22}$ and $sγ_{33}$ are not exists then Set SCOUNT$_1$ = 0, SCOUNT$_2$ = 0 and SCOUNT$_3$ = 0.

**Step48** Multiply the number of occurrence calculated at Step47 with the weights $w_1$, $w_2$ and $w_3$. Now TERM_WEIGHT$_1$ = SCOUNT$_1$ * $w_1$,

```
TERM_WEIGHT₂ = SCOUNT₂ * w₂, TERM_WEIGHT₃ =
SCOUNT₃ * w₃.
Step49  Add  these  term  weights  to
REL_ONT₁_σ, REL_ONT₂_σ and REL_ONT₃_σ. So
updated value of Relevancy of that Web page
(σ)  for  the  different  domains  are
REL_ONT₁_σ  =  REL_ONT₁_σ  +  TERM_WEIGHT₁,
REL_ONT₂_σ  =  REL_ONT₂_σ  +  TERM_WEIGHT₂,
REL_ONT₃_σ = REL_ONT₃_σ + TERM_WEIGHT₃.
Step50 Select the next SynTerms from
SynTables of Ontology₁, Ontology₂ and
Ontology₃ respectively for γ₁, γ₂ and γ₃ and
go to Step47 until all the SynTerms from
SynTables for γ₁, γ₂ and γ₃ are visited.
Step51 Select the next terms for all
domains and weights from weight tables and
go to Step43 until all the terms for all
domains are visited.
Step52 End.
```

In Figure 4 we describe how the above algorithm works to calculate relevance scores. First we take ontology terms for different domains. Then we were finding common terms for minimizing comparison.

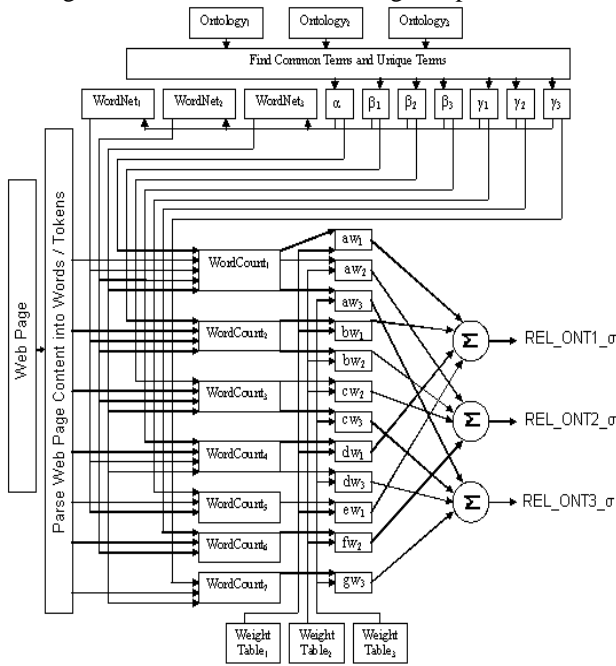

**Fig.4.** Relevance calculation of a Web page

We extracts the terms (α) which belongs to all domains (here we working with three domains). Then find the terms ($\beta_1$, $\beta_2$ and $\beta_3$) from the remaining ontology terms which belongs to any two ontologies i.e. two domains and the remaining terms ($\gamma_1$, $\gamma_2$ and $\gamma_3$) belong to a single domain. All terms have a weight and it varies domain to domain for a single term. Each ontology term has an entry to syntable which contains the synonyms of the ontology terms. Here syntales are $WordNet_1$, $WordNet_2$ and $WordNet_3$ and weight tables are $WeightTable_1$ $WeightTable_2$ and $WeightTable_3$. Now to finding relevance scores we first calculate number of occurrence of ontology terms and corresponding syntable terms. After calculating number of occurrence we multiply with corresponding weight value of the respective domains and finally add all multiplied weights domain wise, we get relevance score of a particular domain.

From the Figure 4 we can see that the relevance scores of the Web page (σ) are $REL\_ONT_1\_\sigma$, $REL\_ONT_2\_\sigma$ and $REL\_ONT_3\_\sigma$ for three domains.

### 5.2 Checking Domain of a Web page

Using ontological knowledge we can find relevant Web pages from the Web. When the crawler finds a new page then it calculate the relevance of the Web page (i.e. it compares the content of the Web page with ontological knowledge). If the calculated relevance is more than a predefined relevance then we called the Web page is of the specific domain. If a Web page overcomes all the relevance limits for all domains then we called the Web page belongs to all domains. If any Web page belongs to any domain we store it in our page repository and also store the relevance scores for further use. For a Web page there are a number of link associated with it, so we need to take special care about the links to make our crawler focus on the specific domain.

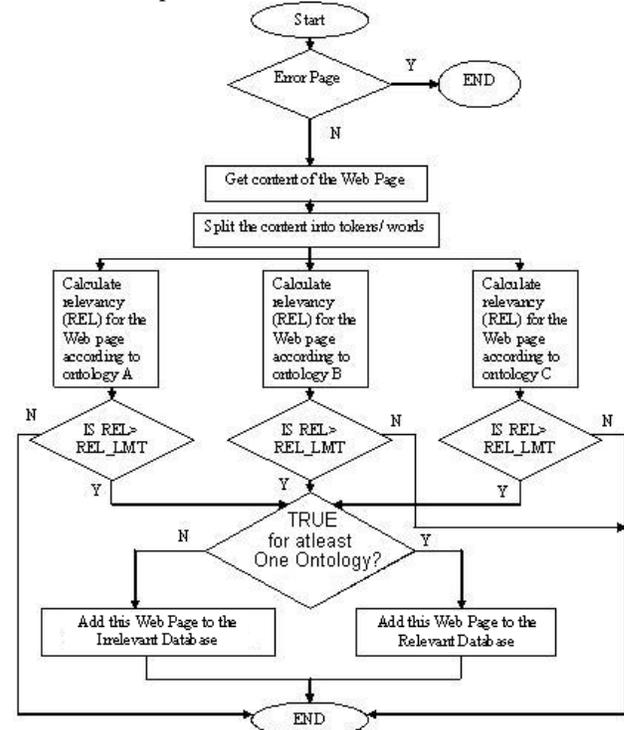

**Fig.5.** Checking Domain of a Web page.

### 5.3 How to Collect Relevant Pages from Irrelevant Links

In our approach we go along the link what are found in domain specific pages. We are not checking link found in the irrelevant pages. If some domain specific pages are partitioned by some irrelevant pages which are not of the specific domain then, the performance of the crawler will degrade. From the Figure 6 we can see that at level 1 there are some irrelevant pages which are discarding domain specific pages at level 2 and 3 from the crawling path. If we can't process those pages then the performance of the crawler will degrade.

As a solution [7] of this problem we take a tolerance limit, this tolerance limit is a very important criterion. When some page is irrelevant then the URLs found in the Web page are stored in a different table we call this table as *IRRE_TABLE*. The *IRRE_TABLE* has two columns URL and Level. We crawl down through those URLs in *IRRE_TABLE* up to the tolerance limit level. If some relevant page found then page is added to the main database. If no relevant pages were found then the URLs are discarded.

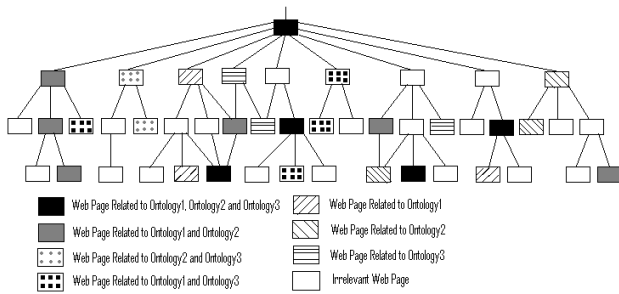

**Fig.6.** Challenge in our approach.

### 5.4 Generation of Graph

In this section we design an algorithm which generates a weighted graph according to their domains. Each domain or ontology represented by a node and these nodes are plot on the 2D plane. Now we considering another node into space that represents common pages i.e. the page belong to all domains. Here we are taking a Web page from page repository and find the domain by relevance value of that page. Each node and each edge in the plane contains a Database and the edges in space are contains a weight. The edge weight between two nodes in the plane contains the common pages for these two domains. The edge weight between two nodes into space contains a numeric value which came from counting number of pages from space node. In space all the edges contain same weight. And the following algorithm generates a weighted graph for all domains.

```
INPUT: A set of Web pages and their
relevance scores for all domains.
OUTPUT: A Weighted Graph.
Step1 Assign node for each Ontology. Here
we assign A, B and C for Ontology₁,
Ontology₂ and Ontology₃ respectively and
assign another node D in space for storing
all domains related pages. Each node and
each edge in the plane contains a Database
and the edges in space are contains
a weight. Initially all the Databases are
blank and all weights in the space edges
are 0.
Step2 Find out the Web pages which are
relevant to only one domain i.e. relevance
score cross the relevance limit for only
one domain and Store the Web pages in the
respective node Database.
Step3 Find out the Web pages which are
relevant to all domains i.e. relevance
score cross the relevance limit for all
domains and Store the Web pages into the
space node (i.e. node D).
Step4 Count number of pages in the space
node and assign the space edge weights by
the count value.
Step5 Find out the Web pages which are
relevant to any two domains i.e. relevance
score cross the relevance limit for any two
domains and Store the Web pages in the
respective edge Database.
Step6 End.
```

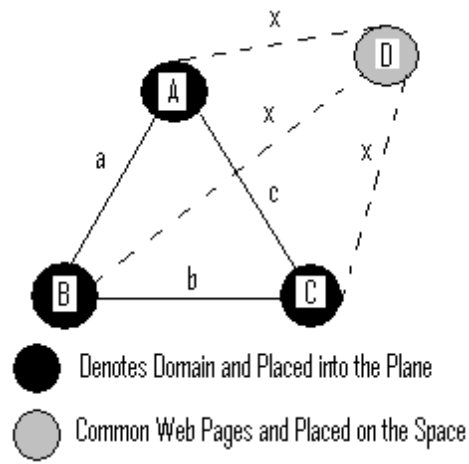

**Fig.7.** Graph representation of Web pages.

In the above graph *A, B* and *C* represent three domains and contain a database of single domain pages. *a, b* and *c* are weights which contains set of web pages. '*a*' contain *A* and *B* domain related pages. '*b*' contain *B* and *C* domain related pages. '*c*' contain *A* and *C* domain related pages. And '*x*' contains number of pages in *D*. node *D* contains pages which belongs to all domains.

### 5.5 User Interface

In Figure 8 shows a part of User Interface in our search engine. Initially *Go* button can't appear in the User Interface. First we put a search query into the *Input String Box* then select domains. After domain selection *Go* button appears on the screen. Here we are working with three domains *Cricket, Football* and *Hockey*. These domains are very closer to each other and our challenges are to find pages from such close domains. Some terms like *Ground, Player, Ball* etc. are applicable to all three domains but some terms are their which are unique to each domain. We are giving strength of those unique terms to find more accurate results. We are using *Check Box* to select domains because if any search string may belongs to all three domains then we select all three domains to find relevant results or if user does not know the search string belongs which domain but user know that the string belongs any of these three domains, in that type of situation user can also select all the domains. Now what activity going on after clicking *Go* button, first parse the search string and then we simply search according to that parsed query on the graph which are generated in section 5.4.

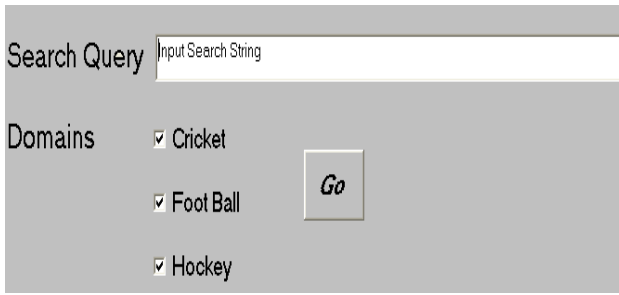

**Fig.8.** A Part of User Interface.

## 6. Performance Analyses

In this section we describe our test settings and describe the performance of our system.

**6.1 Test Settings**
In this section we will describe different parameter settings to run the crawler.

**6.1.1 Seed URLs**
For the crawler to start crawling we provide some seed URLs depending on the Ontologies.

http://www.hindustantimes.com, http://www.cricket-time.com, http://www.sportsofworld.com, http://icc-cricket.yahoo.com, http://www.hockeygiant.com, http://www.whockey.com, http://www.fifa.com, http://www.webindia123.com/sports/hockey/index.htm, http://www.footballtransfers.info, http://www.napit.co.uk, http://www.footballguys.com.

**6.1.2 Syntable**
Synonyms for each Ontology terms are shown in Figure 9, 10 and 11. The Syntables are constructed using different Ontologies. This table contains two columns; one column for Ontology terms and another for synonyms of that term. Here NA defines no such synonyms are present there.

| National Match | Intra state game |
|---|---|
| Not Out | Batting |
| Off Stump | Right side Wicket |
| One day | 50 over match |
| Out | Dismissed |

**Fig.9.** Syntable for Cricket Ontology

| Center | middle |
|---|---|
| Centre Circle | NA |
| Club | Association |
| Corner | area |
| Crowd | mass |

**Fig.10.** Syntable for Foot-Ball Ontology

| Defender | protector |
|---|---|
| Draw | NA |
| Elbow Pads | NA |
| EQUIPMENTS | Apparatus |
| Field Hockey | NA |

**Fig.11.** Syntable for Hockey Ontology

**6.1.3 Weight Table**
Weight for each Ontology terms is shown in Figure 12, 13 and 14. The weight tables are constructed using different Ontologies. This table contains two columns; one column for Ontology terms and another for weight of that term.

| Not Out | 0.8 |
|---|---|
| Off Stump | 0.8 |
| Out | 0.6 |
| One day | 0.4 |
| National Match | 0.1 |

**Fig.12.** Weight table for Cricket Ontology

| Free kick | 0.8 |
|---|---|
| Centre Circle | 0.4 |
| Corner | 0.4 |
| Center | 0.2 |
| Crowd | 0.1 |

**Fig.13.** Weight table for Foot-Ball Ontology

| Field Hockey | 0.9 |
|---|---|
| Hockey Stick | 0.9 |
| Elbow Pads | 0.6 |
| Defender | 0.2 |
| Draw | 0.1 |

**Fig.14.** Weight table for Hockey Ontology

### 6.2 Test Results

In this section we have shown some test results through graph plot.

### 6.2.1 Performance of multiple domains crawling over single domain crawling

From the Figure 15 we can see that, single domain specific crawler crawling time is more than the multiple domains specific crawler crawling time. When we work through large number of Web pages in single domain specific crawler, most of the Web pages are irrelevant and we discard those pages but in multiple domains specific crawler, most of the pages does not irrelevant page, it belongs to any one domain and if these domains are match with our domains then our crawler performance increase.

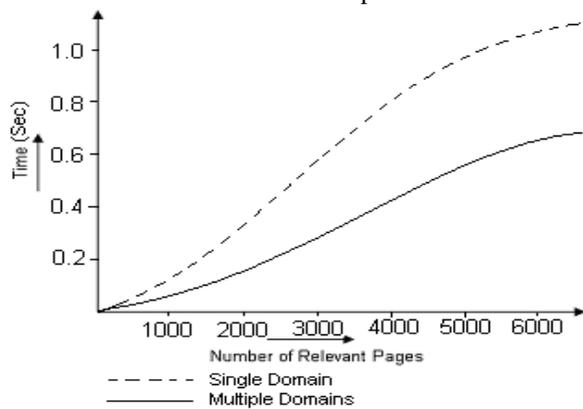

**Fig.15.** Time taken in Single Domain Crawling and Multiple Domains crawling.

### 6.2.2 Page Distribution in Different Domains

In Figure 16 we have shown page distribution of each domain.

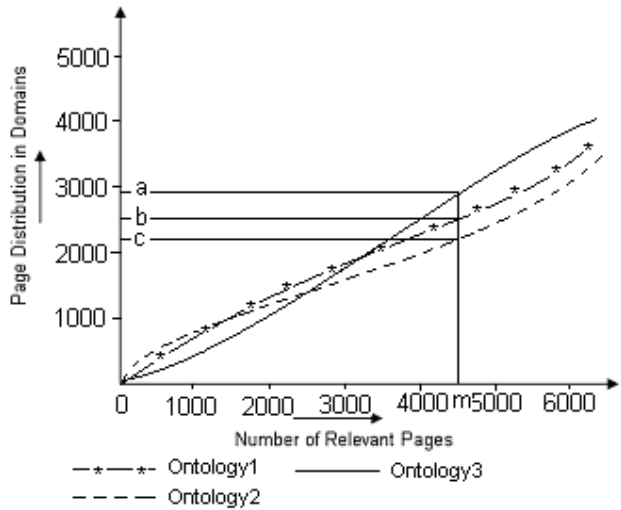

**Fig.16.** Page Distribution in Domain wise.

From the figure we conclude that one page must be belongs to more than one domain. Here *m* is number of relevant pages and *b, c* and *a* number of relevant pages belongs to domain *1, 2* and *3* respectively and m always less than equal to (*a+b+c*).

## 7. Conclusions

Web searchers faced major problems by imprecise and irrelevant results, especially with the continued expansion of the Web. For this we incorporate domain specific concept for crawling Web pages from WWW.

In our experiment, we have developed a prototype that uses multiple ontologies to perform multiple domains specific crawling. The prototype uses information of a specified domains are kept in structure way into ontology to guide the crawler in its search for Web pages that are relevant to the topics specified in ontologies.

Firstly, our approach has been able to successfully eliminate the problem of irrelevant results which is one of the main problems encountered by the users of a regular search engine. By searching domain specific Web pages the search engine effectively fetches the exact information.

Secondly, by producing exact information as the result, the search engine eliminates the need to go through numerous results as in case of a regular search engine.

Thirdly, effectiveness of multiple domains specified search engine better than other search engines.

Finally, our design although based on three domains, is highly scalable and can be easily adopted by other enterprises as their site search tool. This would only require the enterprise to feed in the relevance limit, weight tables based on the ontology of the different domains.